# Landau-Zener evolution under weak measurement: Manifestation of the Zeno effect under diabatic and adiabatic measurement protocols

Anna Novelli,[1,2] Wolfgang Belzig[1] and Abraham Nitzan[2]

[1] Department of Physics, University of Konstanz, D-78457 Konstanz, Germany
[2] School of Chemistry, Tel Aviv University, Tel Aviv 69978, Israel

**Abstract**

The time evolution and the asymptotic outcome of a Landau-Zener-Stueckelberg-Majorana (LZ) process under continuous weak non-selective measurement is analyzed. We compare two measurement protocols in which the populations of either the adiabatic or the non-adiabatic levels are (continuously and weakly) monitored. The weak measurement formalism, described using a Gaussian Kraus operator, leads to a time evolution characterized by a Markovian dephasing process, which, in the non-adiabatic measurement protocol is similar to earlier studies of LZ dynamics in a dephasing environment. Casting the problem in the language of measurement theory makes it possible for us to compare diabatic and adiabatic measurement scenarios, to consider engineered dephasing as a control device and to examine the manifestation of the Zeno effect under the different measurement protocols. In particular, under measurement of the non-adiabatic populations, the Zeno effect is manifested not as a freezing of the measured system in its initial state, but rather as an approach to equal asymptotic populations of the two diabatic states. This behavior can be traced to the way by which the weak measurement formalism behaves in the strong measurement limit, with a built-in relationship between measurement time and strength.

## 1. Introduction

The quantum Zeno effect - the suppression of time evolution between discrete quantum states under frequent repeated measurement - is well understood as a consequence of the general theory of the time evolution of a quantum system that interacts with its environment. In the simplest manifestation of this effect, interstate transitions in an interacting two-level system are shown to slow down under repeated interrogation of the level populations. When discussed in the framework of measurement theory, this behavior reflects the wavefunction collapse upon determination of the quantum state. In the more general weak measurement theory the effect on the system of a continuous weak measurement can be cast as a decoherence process whose rate reflects the measurement weakness. Indeed, the time evolution of a quantum system interacting with its environment is usually discussed without making connection to an underlying measurement process. Still, it is sometimes useful to make this connection for its conceptual value as well as its experimental implication. To elaborate, consider a two level system that represents an electron tunneling between the two minima of a double-well potential and assume that temperature is low enough so that only the two lowest electronic states in this potential can be occupied. We may choose to measure the charge state of one of the wells using a nearby point contact device or we may devise a spectroscopic tool that monitors the population of the true system ground state (a linear combination of the two states localized in each wells). These different measurement protocols have different effects on the system dynamics and their consideration may provide insight on the interrelationship between measurement, decoherence and quantum time evolution.

In this paper we consider the effect of continuous weak measurement on the time evolution of a Landau-Zener-Stueckelberg-Majorana[1-4] process. In the so-called diabatic representation, the Hamiltonian of this well-known model describes two coupled levels with time dependent energy spacing

$$\hat{H}^{dia}(t) = \begin{pmatrix} ut & V \\ V & -ut \end{pmatrix} \qquad (1)$$

The superscript *dia* indicates that this Hamiltonian is represented in the so called diabatic basis. Denoting the general time dependent solution of the Schrödinger equation in this representation by $\Psi(t) = c_1(t)\begin{pmatrix}1\\0\end{pmatrix} + c_2(t)\begin{pmatrix}0\\1\end{pmatrix}$, the aim is to find the wavefunction at time $t \to \infty$, given that in the distant past it was in state 1 (say). Specifically, we are interested in the probabilities $|c_1(t)|^2$ and $|c_2(t)|^2 = 1 - |c_1(t)|^2$ for $t \to \infty$ given that $|c_1(t)|^2 = 1$ at $t \to -\infty$. The Landau-Zener result is

$$|c_1(t \to \infty)| = \exp\left(-\frac{\pi V^2}{u}\right) \qquad (2)$$

Alternatively we may represent the problem in the time dependent adiabatic basis, $(\psi_a(t), \psi_b(t))$ that diagonalizes the instantaneous Hamiltonian at any point in time with the corresponding eigenvalues $E_a(t), E_b(t) = \pm\sqrt{u^2 t^2 + V^2}$, and describe the time evolution in terms of this basis $\Psi(t) = c_a(t)\psi_a(t) + c_b(t)\psi_b(t)$. Obviously, the asymptotic $(t \to \pm\infty)$ values of $c_{a,b}(t)$ and $c_{1,2}(t)$ are identical. There is a large body of work that address the effect of coupling to an external thermal environment on this evolution,[5-14] [15, 16] [17-21] including the possibility of externally affected control.[22]

As pointed out above, and further demonstrated below, the effect of continuous weak quantum measurement on a system can be cast as a dephasing process. As such, its description is strongly related to the above studies. Indeed, some of these works address detailed properties of the external bath, including its temperature, that are not usually included in standard descriptions of measurement. On the other hand, discussing this time evolution as a consequence of a measurement process can highlight issues that are not naturally considered otherwise. For example, most of the works cited above focus on a particular model of system bath coupling, where the diagonal matrix elements of the Hamiltonian in the diabatic representation are randomly modulated or otherwise linearly coupled to a harmonic thermal bath. In the framework of measurement theory it is natural to define first the nature of the measurement. In particular, we may consider monitoring the populations of the adiabatic levels or of the diabatic levels, with possibly different consequences on the ensuing time evolution. This distinction may come up in specific experimental situations. For example, in many applications, the two diabatic states represent electron localization on different sites in the system (in which case the coupling $V$ in Eq. (1) is the interaction responsible for electron transfer between the two sites). Measurement of the corresponding population may be done by monitoring the charge on one of these sites using a nearby quantum point contact whose transmission (hence the corresponding monitored current) is sensitive to this charge, see e.g., Ref. [23]. On the other hand, it is possible to monitor the instantaneous population of the (adiabatic) electronic eigenstates of a system, as was done in the possibly first experimental demonstration of the quantum Zeno effect [24] [25] (see also Ref. [28] for a recent demonstration of such measurement).

In this paper we discuss the realization of the Zeno effect under these two types of measurement. In the next Section we briefly review the theory of continuous weak measurement in the Kraus operator formalism[29, 30] and discuss the time evolution of a system characterized by the Hamiltonian (1) under continuous weak measurements of its adiabatic or diabatic state populations. Numerical results for the corresponding time evolutions are presented in Section 3, showing the different manifestations of the Zeno effect in the strong measurement limit of these two schemes. Section 4 concludes.

## 2. LZ dynamics under continuous measurement

*Weak measurements and Kraus operators.* A generalized quantum measurement[30, 31] is described by a set of measurement operators $\{\hat{K}_a(\hat{A})\}$ fulfilling the completeness relation $\int da \hat{K}_a^\dagger(\hat{A}) \hat{K}_a(\hat{A}) = \hat{1}$ in the case of a continuous spectrum of measurement outcomes. We are interested in a concrete form of a measurement operator, which is able to describe a fuzzy measurement process. We expect that this operator[32, 33] depends on a parameter $\bar{\lambda}$, which defines the strength of measurement and is hence related to its resolution. This parameter should provide the ability to interpolate continuously between the hard projective measurement and a fuzzy measurement with very few impact on the system. Intuitively we expect that it is more probable that an actual eigenvalue of $\hat{A}$ lies close to the measured value $a$ and that the probability to be the actual value then decreases smoothly by growth of $|A-a|$. Hence, the measurement operator is approximated by a Gaussian form with a single parameter $\bar{\lambda}$

$$\hat{K}_a(\hat{A}) \equiv \left(\frac{2\bar{\lambda}}{\pi}\right)^{1/4} \exp\left[-\bar{\lambda}(a-\hat{A})^2\right] \qquad (3)$$

It is easy to check that the completeness relation is satisfied. Furthermore it is clear that for $\bar{\lambda} \to \infty$ we obtain an operator which describes a strong, exact measurement as the Gaussian becomes very narrow and peaked for the eigenvalues of $\hat{A}$, while $\bar{\lambda} \to 0$ corresponds to a very weak measurement with fuzzy observations and the Krausoperator become almost $\hat{1}$.

The probability density $\mathcal{P}(a)$ to obtain a result $a$ of a measurement is in general given by

$$\mathcal{P}(a) = \text{Tr}\left[\hat{K}_a(\hat{A})\hat{\rho}\hat{K}_a^\dagger(\hat{A})\right]. \tag{4}$$

The normalized density matrix after such a measurement is

$$\hat{\rho}_{\text{after},a} = \hat{K}_a(\hat{A})\hat{\rho}\hat{K}_a^\dagger(\hat{A})/\mathcal{P}(a) \tag{5}$$

The density matrix formalism provides the ability to treat nonselective measurements. We perform a measurement on a system and the output is registered but not used. Accordingly, we obtain for the nonselective post measurement density matrix:

$$\hat{\rho}_{\text{after}}^{\text{nonsel}} = \int da\, \hat{\rho}_{\text{after},a} = \int da\, \hat{K}_a(\hat{A})\hat{\rho}\hat{K}_a^\dagger(\hat{A}) \tag{6}$$

We have to sum over the unnormalized selective density matrix to conserve the normalization of $\hat{\rho}_{\text{after}}^{\text{nonsel}}$. This is the same as the sum over all normalized selective matrices weighted with the probability $\mathcal{P}(a)$.

*Continuous weak measurement.* The previous definition of a measurement can be easily generalized to a continuous measurement. Naivly, continuous projective measurement would cause a total suppression of the dynamics analogously to the quantum Zeno effect [31]) due to the continuous collapse of the wave function into an eigenstate. Alternatively we can consider continuous weak measurements which provide less information but do not disturb the system to such an extent. The question is whether it is possible to obtain sufficient information with continuous weak measurement while leaving the system as undisturbed as needed.

A general description of the time evolution of a system under continuous weak measurement can be derived by approximating this evolution through repeated instantaneous measurements at consecutive time instants $t_i$, equally separated by small time steps $\Delta t$.[32] A single measurement of an observable $\hat{A}$ that yields the outcome $a_i$ corresponds to the application on the system of a measurement operator $\hat{K}_{a_i}(\hat{A})$, in our case the before derived Gaussian Kraus operator. During the time $\Delta t$ between two successive measurements at times $t_i$ and $t_{i+1}$, the system evolves freely according to Schrödinger's equation, which is expressed by the unitary time evolution operator $\hat{U}(t_{i+1}, t_i)$:

$$\hat{U}(t_{i+1},t_i) = \mathcal{T} \exp\left[-\frac{i}{\hbar}\int_{t_i}^{t_{i+1}} dt \hat{H}(t)\right] \overset{\Delta t H \ll 1}{\approx} \exp\left[-\frac{i}{\hbar}\hat{H}(t_i)\Delta t\right]. \tag{7}$$

The time evolution of the system after time $t = (N+1)\Delta t + t_0$ after a sequence of $N$ weak measurements separated by intervals of free time evolution is then given by:

$$\hat{U}(t,t_N)\hat{K}_{a_N}\hat{U}(t_N,t_{N-1})...\hat{U}(t_2,t_1)\hat{K}_{a_1}\hat{U}(t_1,t_0) \tag{8}$$

Furthermore, it is assumed that the measurement strength is inversely proportional to its frequency[32]

$$\bar{\lambda} = \lambda \Delta t \tag{9}$$

with constant $\lambda$. Then we have the following form of the Kraus operator:

$$\hat{K}_{a_i}^{\lambda}(\hat{A}_i) \equiv \left(\frac{2\lambda \Delta t}{\pi}\right)^{1/4} \exp\left[-\lambda(a_i - \hat{A}_i)^2 \Delta t\right]. \tag{10}$$

And in the continuum limit $\Delta t \to 0$ ($N \to \infty$), we obtain up to a normalization factor

$$\mathcal{T} \exp\left[\int_{t_0}^{t} d\tau \left(-\frac{i}{\hbar}\hat{H}(\tau) - \lambda(a(\tau) - \hat{A}(\tau))^2\right)\right] \equiv \hat{K}_{[a]}^{\lambda}(t) \tag{11}$$

The discrete results $a_i$ become a function $a(t)$; the operator $\hat{K}_{[a]}^{\lambda}(t)$ describes the full time evolution of a system under continuous weak measurement with this outcome $a(t)$. Note that the observable $\hat{A}(t)$ can in general change with time (and the time-dependence here is not signaling the Heisenberg picture). For a nonselective measurement this time evolution has to be applied to the density matrix and then integrated over all intermediate measurement outcomes.

In Ref. [32] it was shown (see also [31]) that the time evolution of a density matrix $\hat{\rho}_{[a]}(t) = \hat{K}_{[a]}^{\lambda}(t)\rho\hat{K}_{[a]}^{\lambda}(t)$ undergoing a nonselective weak measurement of the observable $\hat{A}(t)$ is given by a Lindblad master equation:

$$\frac{d}{dt}\hat{\rho}(t) = -\frac{i}{\hbar}\left[\hat{H}(t), \hat{\rho}(t)\right] - \frac{\lambda}{2}\left[\hat{A}(t), \left[\hat{A}(t), \hat{\rho}(t)\right]\right] \tag{12}$$

In applying this general result to the LZ evolution under continuous measurement we can choose to monitor populations in the diabatic states or in the adiabatic states. The former measurement mode can be accomplished by choosing

$$\hat{A} = \hat{\sigma}_z \equiv \begin{pmatrix} 1 & 0 \\ 0 & -1 \end{pmatrix} \tag{13}$$

In Eq. (12). This leads to

$$\begin{aligned}
\frac{d}{dt}\rho_{11}^{dia} &= -iV(\rho_{21}^{dia} - \rho_{12}^{dia}) \\
\frac{d}{dt}\rho_{12}^{dia} &= -i2ut\rho_{12}^{dia} - iV(\rho_{22}^{dia} - \rho_{11}^{dia}) - 2\lambda\rho_{12}^{dia} \\
\frac{d}{dt}\rho_{21}^{dia} &= i2ut\rho_{21}^{dia} - iV(\rho_{11}^{dia} - \rho_{22}^{dia}) - 2\lambda\rho_{21}^{dia} \\
\frac{d}{dt}\rho_{22}^{dia} &= iV(\rho_{21}^{dia} - \rho_{12}^{dia})
\end{aligned} \tag{14}$$

For the other possibility, continuous measurement of the adiabatic populations, the measurement observable is the transformed operator

$$\hat{A}^{adi}(t) = \hat{U}^{-1}(t)\hat{\sigma}_z\hat{U}(t) \tag{15}$$

where $\hat{U}(t)$ is the unitary trasformation that diagonalizes the instantaneous Hamiltonian (1)

$$\hat{H}^{adi} = \hat{U}(t)H^{dia}\hat{U}^{-1}(t) = \begin{pmatrix} \sqrt{u^2t^2 + V^2} & 0 \\ 0 & -\sqrt{u^2t^2 + V^2} \end{pmatrix} \tag{16}$$

Alternatively (and equivalently) we can represent the dynamics in the adiabatic basis, where $\psi^{adi} = U\psi^{dia}$ evolves according to

$$\frac{d\psi^{adi}}{dt} = -i\left(\hat{H}^{adi} - i\hat{U}(t)\frac{d\hat{U}(t)^{-1}}{dt}\right)\psi^{adi} \tag{17}$$

The evolution of the density operator is similarly modified. Eq. (12) becomes

$$\frac{d\hat{\rho}^{adi}}{dt} = -i\left[\left(\hat{H}^{adi} - i\hat{U}(t)\frac{d\hat{U}(t)^{-1}}{dt}\right), \hat{\rho}^{adi}\right] - \frac{\lambda}{2}\left[\hat{A}(t), \left[\hat{A}(t), \hat{\rho}^{adi}\right]\right] \tag{18}$$

Where, in this representation, the operator $\hat{A}$ that measures populations in the adiabatic basis is again given by Eq. (13). This leads to

$$\frac{d}{dt}\rho_{11}^{adi} = \left[\hat{M}(t), \hat{\rho}^{adi}\right]_{11}$$
$$\frac{d}{dt}\rho_{12}^{adi} = -2i\sqrt{u^2t^2 + V^2}\rho_{12}^{adi} + \left[\hat{M}(t), \hat{\rho}^{adi}\right]_{12} - 2\lambda\rho_{12}^{adi}$$
$$\frac{d}{dt}\rho_{21}^{adi} = 2i\sqrt{u^2t^2 + V^2}\rho_{21}^{adi} + \left[\hat{M}(t), \hat{\rho}^{adi}\right]_{21} - 2\lambda\rho_{21}^{adi} \quad (19)$$
$$\frac{d}{dt}\rho_{22}^{adi} = \left[\hat{M}(t), \hat{\rho}^{adi}\right]_{22}$$

Where

$$\hat{M}(t) == \hat{U}(t)\frac{d\hat{U}^{-1}(t)}{dt} \quad (20)$$

The explicit form of the transformation matrix $U$ is given by

$$\hat{U}(\tilde{t}) = \begin{pmatrix} \cos\frac{\theta(t)}{2} & \sin\frac{\theta(t)}{2} \\ -\sin\frac{\theta(t)}{2} & \cos\frac{\theta(t)}{2} \end{pmatrix} \quad (21)$$

With $\tan\theta(t) = V/ut$. Eqs. (13)-(21) are used to obtain the results presented and discussed next.

## 3. Results and Discussion

It is convenient to display the results in terms of dimensionless parameters. Define

$$\tilde{t} = \frac{u}{V}t, \qquad z = \frac{V^2}{u}, \qquad \tilde{\lambda} = \frac{V}{u}\lambda = \frac{z}{V}\lambda \quad (22)$$

In terms of these variables Eqs. (14) become

$$\frac{d}{d\tilde{t}}\rho_{11}^{dia} = -iz(\rho_{21}^{dia} - \rho_{12}^{dia}) \quad (23a)$$

$$\frac{d}{d\tilde{t}}\rho_{12}^{dia} = -2iz\tilde{t}\rho_{12}^{dia} - iz(\rho_{22}^{dia} - \rho_{11}^{dia}) - 2\tilde{\lambda}\rho_{12}^{dia} \quad (23b)$$

$$\frac{d}{d\tilde{t}}\rho_{21}^{dia} = 2iz\tilde{t}\rho_{21}^{dia} + iz(\rho_{22}^{dia} - \rho_{11}^{dia}) - 2\tilde{\lambda}\rho_{21}^{dia} \quad (23c)$$

$$\frac{d}{d\tilde{t}}\rho_{22}^{dia} = iz(\rho_{21}^{dia} - \rho_{12}^{dia}) \quad (23d)$$

while Eqs. (19) take the form

$$\frac{d}{d\tilde{t}}\rho_{11}^{adi} = \left[\hat{\tilde{M}}(\tilde{t}), \hat{\rho}^{adi}\right]_{11} \tag{24a}$$

$$\frac{d}{d\tilde{t}}\rho_{12}^{adi} = -2iz\sqrt{\tilde{t}^2+1}\rho_{12}^{adi} + \left[\hat{\tilde{M}}(\tilde{t}), \hat{\rho}^{adi}\right]_{12} - 2\tilde{\lambda}\rho_{12}^{adi} \tag{24b}$$

$$\frac{d}{d\tilde{t}}\rho_{21}^{adi} = 2iz\sqrt{\tilde{t}^2+1}\rho_{21}^{adi} + \left[\hat{\tilde{M}}(\tilde{t}), \hat{\rho}^{adi}\right]_{21} - 2\tilde{\lambda}\rho_{21}^{adi} \tag{24c}$$

$$\frac{d}{d\tilde{t}}\rho_{22}^{adi} = \left[\hat{\tilde{M}}(\tilde{t}), \hat{\rho}^{adi}\right]_{22} \tag{24d}$$

with $\hat{\tilde{M}}(\tilde{t}) = \hat{U}(\tilde{t})d\hat{U}^{-1}(\tilde{t})/d\tilde{t}$. $\hat{U}(\tilde{t})$ is given by Eqs. (21) with $ut/V$ replaced by $\tilde{t}$ everywhere. These equations are solved using the 4$^{th}$ order Runge-Kutta algorithm to give the results displayed below. The timestep size was of order $\Delta\tilde{t}$ = 0.002-0.01, chosen so as to insure convergence in the whole range of parameters.

Consider first measurement in the adiabatic basis, where taking $\hat{A} = \hat{\sigma}_z$ to be diagonal in in the adiabatic basis implies that the measurement is aimed to monitor the populations $\rho_{11}^{adi} = 1 - \rho_{22}^{adi}$ of the adiabatic state. Figures 1a and b show the time evolution of these populations, starting from $\rho_{22}^{adi} = 1$ at the distant past, as the system goes through the avoided crossing at $t = 0$, for different values of the Landau-Zener (LZ) parameter $\tilde{z} = 0.05$ and $\tilde{z} = 0.5$, respectively, and different strengths of continuous time measurement $\tilde{\lambda}$. Comparing the no-measurement ($\tilde{\lambda} = 0$) results in the two cases we see the well-known characteristics of these time evolutions: (a) the increased adiabatic nature of the evolution (where the system stays in the initial adiabatic state with larger probability for larger z (larger non-adiabatic coupling or smaller speed) and (b) the small oscillations between the level populations in the neighborhood of the avoided crossing. As $\tilde{\lambda}$ increases both features change in an expected way: the quantum Zeno effect is manifested in the decreasing transition probability between the two states, that is, the evolution becomes more adiabatic. Also, the oscillations are washed out, expressing the phase-destroying nature of the measurement process. It is interesting to note that, counterintuitively, for large $\tilde{z}$ (that is, close to the adiabatic limit) the dependence of the asymptotic population on $\tilde{\lambda}$ is non-monotonous: as $\tilde{\lambda}$ increases from zero the adiabaticity of the processes initially *decreases* before showing the expected increase (see Fig. 1c). This behavior probably results from the fact that in addition to slowing down the dynamics associated with the measured population, strong measurement also causes an effective level broadening and consequently an increase in the transition probability. The second effect is more pronounced in the adiabatic (large $\tilde{z}$) limit when the system evolution is essentially adiabatic even without measurement so imposing measurement can only decrease the adiabatic population.

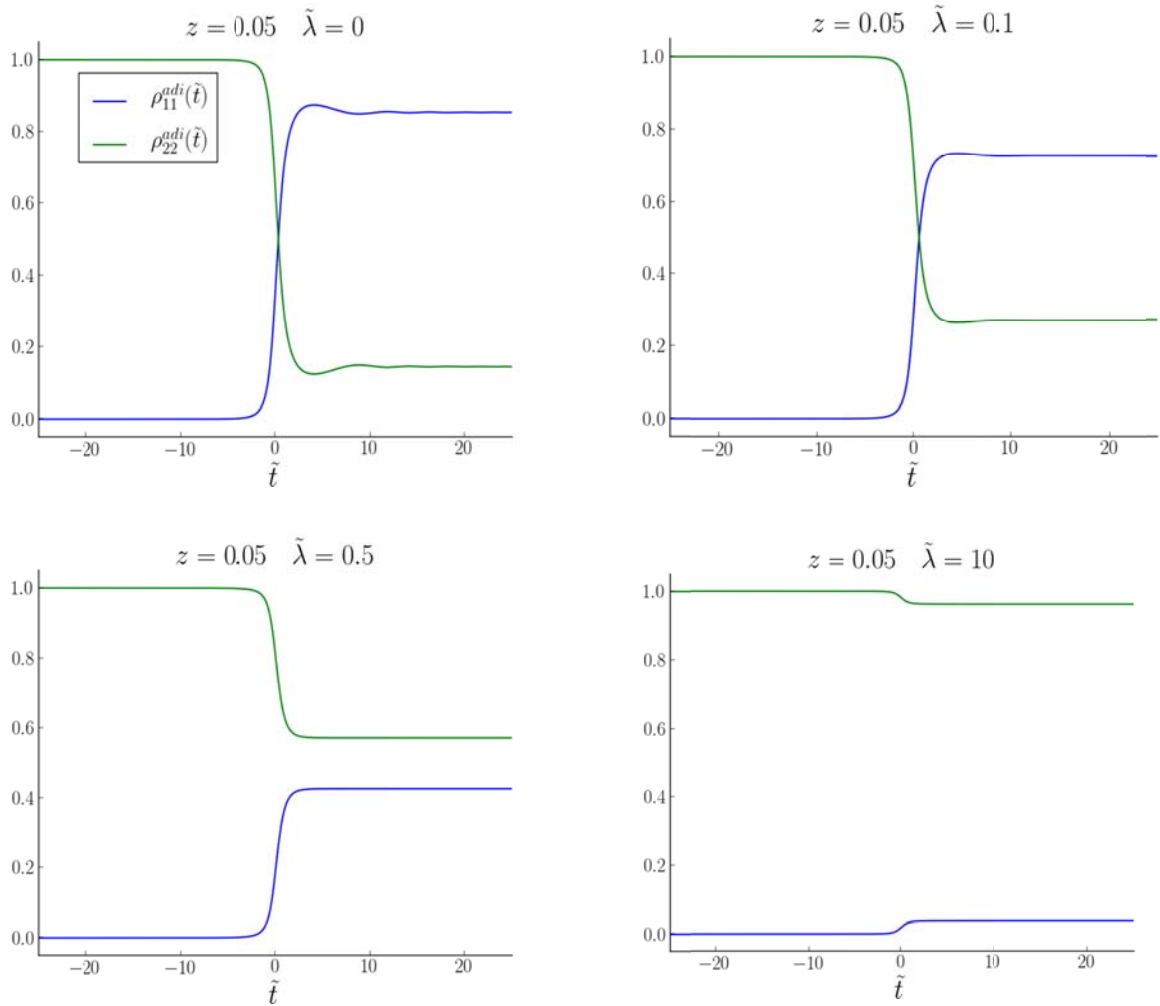

**Fig. 1a**

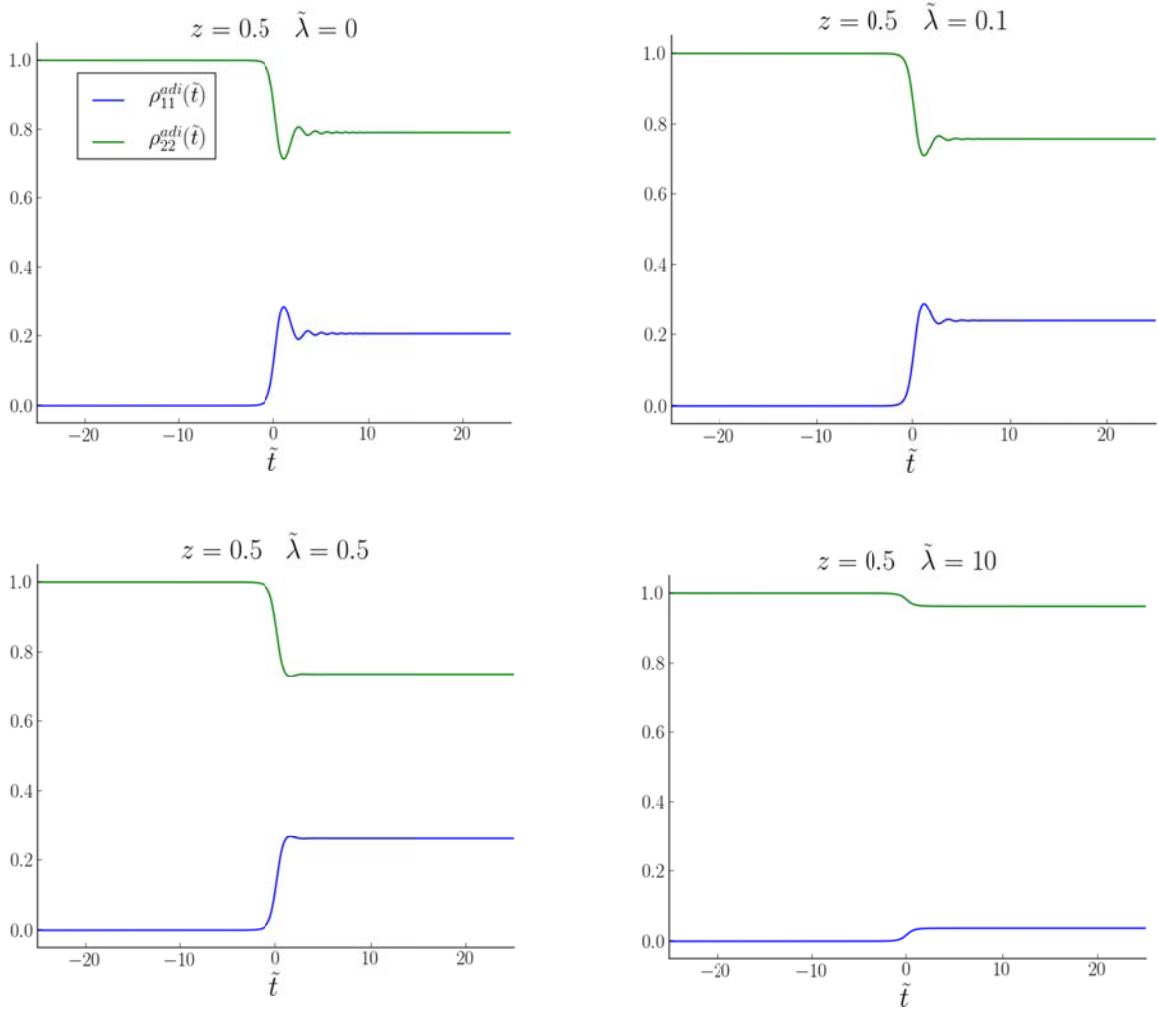

**Fig. 1b**

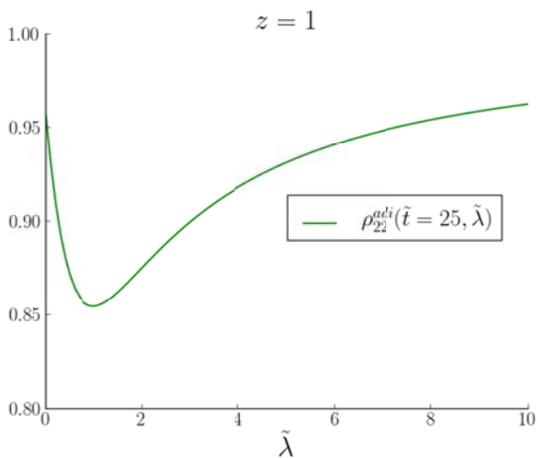

**Fig. 1c**

Fig. 1. The effect of continuous measurement of the adiabatic populations on the LZ transition. Figures 1a and 1b show the time evolution of the populations of the adiabatic states for different values of the LZ parameter, $\tilde{z}$, and for different strengths of measurements, $\tilde{\lambda}$. Fig. 1c shows the asymptotic $(t \to \infty)$ population of the initially populated adiabatic state as function of the measurement strength $\tilde{\lambda}$.

Next consider the same LZ process accompanied by a continuous time measurement of the populations of the diabatic states. Figures 2a-c show respectively results obtained for the nearly non-adiabatic limit, z=0.05, an intermediate case, z=0.5 and the practically adiabatic limit, z=5. In correspondence with the measurement, the displayed populations are those of the diabatic states.[34] Again we observe the typical inter-level interference evidenced by the oscillatory populations near the diabatic crossing that are washed away with increasing measurement strength.[35] More interesting is the way in which the Zeno effect is manifested in the weak and strong measurement regimes as best seen in the z=0.5 results. In the absence of measurement this evolution is fairly adiabatic, and the population of the initially populated diabatic level goes from 1 to ~ 0.2 as the system evolves across the avoided crossing. As $\tilde{\lambda}$ increases from zero the measurement affects an increased non-adiabatic character of the time evolution – an increased probability to remain in the initial non-adiabatic level. However, as $\tilde{\lambda}$ increases further (stronger measurement) this probability assumes the asymptotic value of 0.5. Further increase in $\tilde{\lambda}$ does not change this asymptotic limit, however the typical Zeno behavior is seen in the slowing down of the approach to this limit.

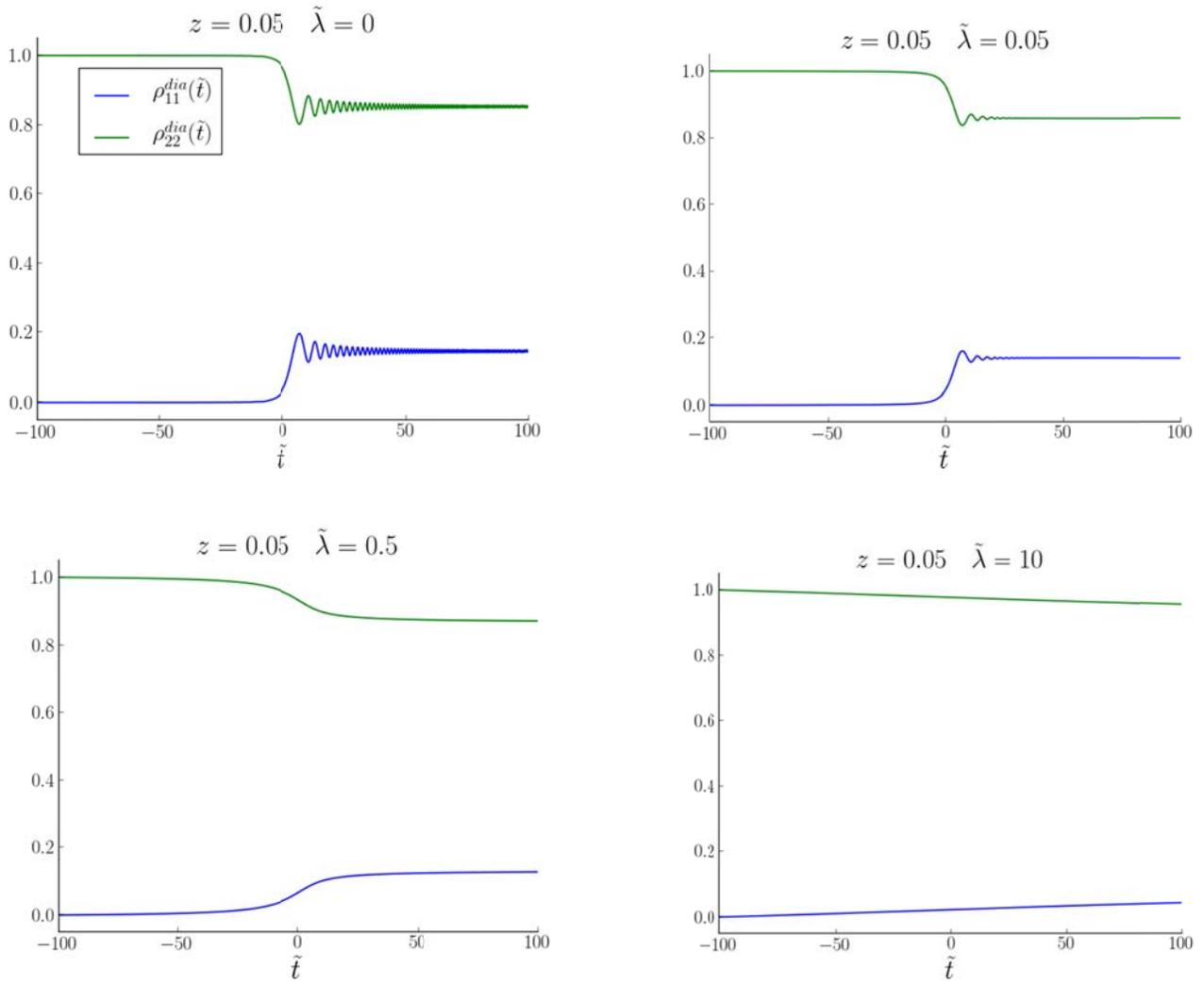

FIG. 2a

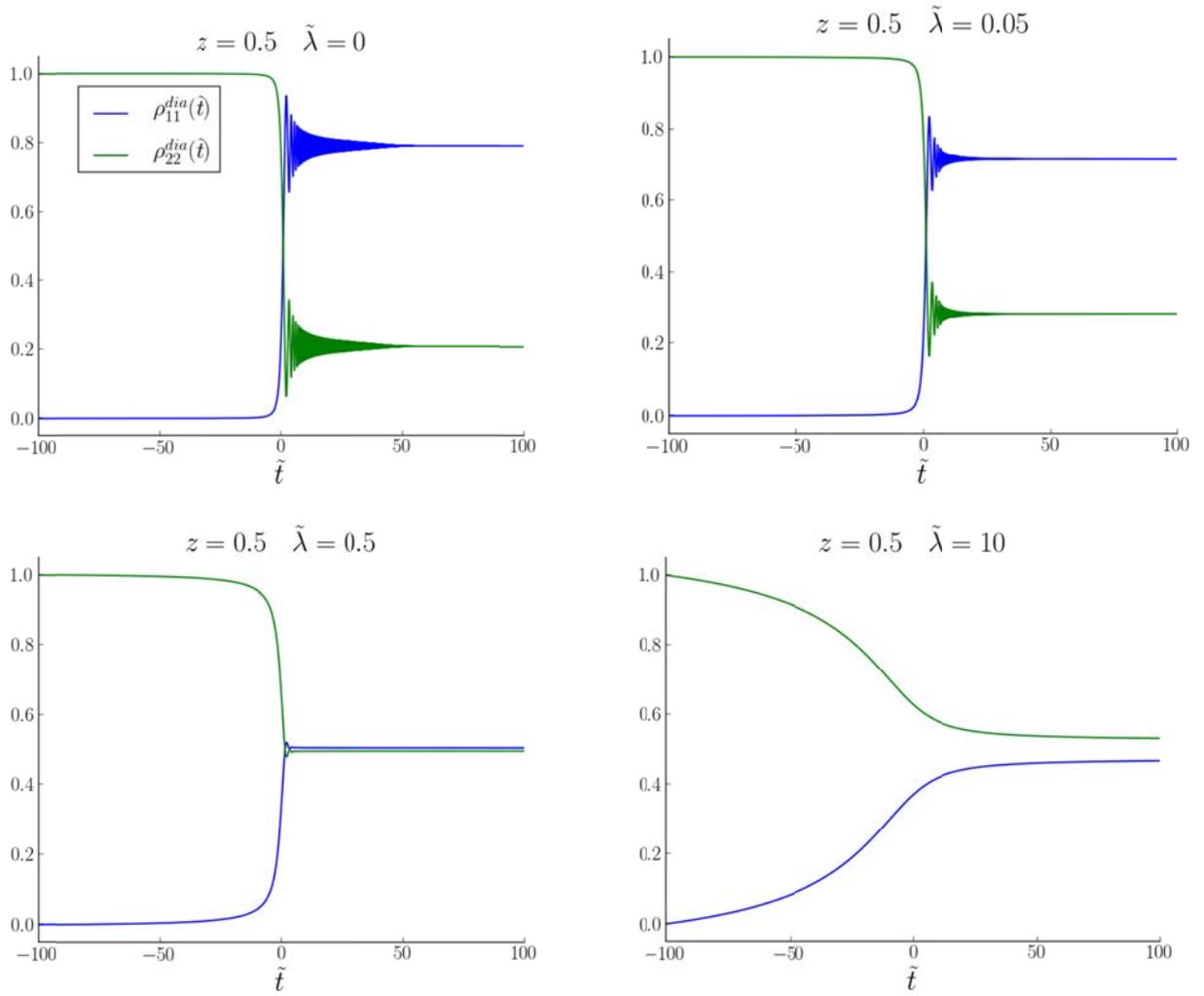

FIG 2b.

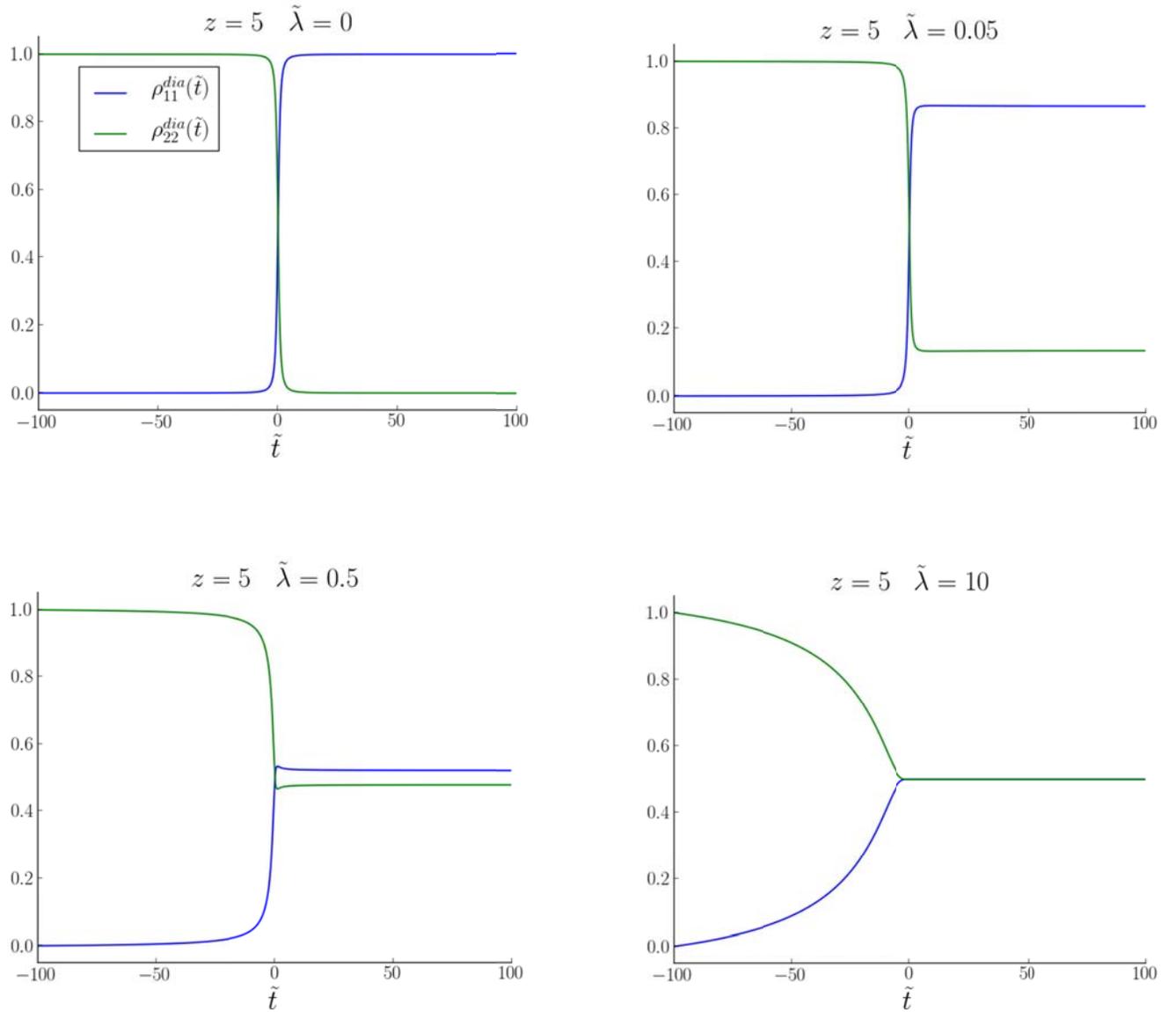

Fig. 2C

Fig. 2. The effect of continuous measurement of the diabatic populations on the LZ transition. Shown are the time evolution of the diabatic states populations for different values of the LZ parameter, $\tilde{z}$, and for different strengths of measurements, $\tilde{\lambda}$. Note that the horizontal timescale here spans a larger range, $-100 < \tilde{t} < 100$, than in Fig. 1 so as to accommodate the expanded evolution range seen in the strong measurement limit.

This observation is similar those made in studies of LZ dynamics under strong dephasing.[5-7, 10, 11] To rationalize it in the present context we start by seeking a solution of Eqs. (23b) in the form

$$\rho_{12}^{dia}(t) = \bar{\rho}_{12}^{dia}(t)\exp\left(-iz\tilde{t}^2 - 2\lambda\tilde{t}\right) \tag{25}$$

Using this in Eq. (23b) and assuming that $\rho_{12}^{dia}(t=0) = 0$ leads to

$$\rho_{12}^{dia}(\tilde{t}) = -iz\int_0^{\tilde{t}} d\tilde{t}\,'\Delta\rho^{dia}(\tilde{t}\,')e^{-iz(\tilde{t}^2-\tilde{t}'^2)-2\tilde{\lambda}(\tilde{t}-\tilde{t}\,')};\quad \Delta\rho(\tilde{t}) = \rho_{22}(\tilde{t}) - \rho_{11}(\tilde{t}) \tag{26}$$

If $\lambda \gg z\tilde{t}$ we may assume that $\Delta\rho$ does not vary appreciably during the lifetime of the integrand in (26), hence

$$\rho_{12}^{dia}(\tilde{t}) \cong -iz\Delta\rho^{dia}(\tilde{t})\int_0^{\tilde{t}} d\tilde{t}\,'e^{-iz(\tilde{t}^2-\tilde{t}'^2)-2\tilde{\lambda}(\tilde{t}-\tilde{t}\,')} \tag{27}$$

Also, in this limit we can approximate $(t^2 - t'^2) \approx 2t(t-t')$. This leads to

$$\rho_{12}^{dia}(\tilde{t}) = \frac{-iz\Delta\rho^{dia}(\tilde{t})}{2izt + 2\tilde{\lambda}}\left(1 - e^{-2iz\tilde{t}^2 - 2\tilde{\lambda}\tilde{t}}\right) \xrightarrow{\tilde{t}\to\infty} \frac{-iz\Delta\rho^{dia}(\tilde{t})}{2iz\tilde{t} + 2\tilde{\lambda}} \tag{28}$$

Using (28) and its complex conjugate in (c.f. (23a,d)) $(d/d\tilde{t})\Delta\rho^{dia} = iz(\rho_{21}^{dia} - \rho_{12}^{dia})$ leads to

$$(d/d\tilde{t})\Delta\rho^{dia} = -\frac{4z^2\tilde{\lambda}}{(2\tilde{\lambda})^2 + (2z\tilde{t})^2}\Delta\rho^{dia} = -\frac{z^2}{\tilde{\lambda}}\Delta\rho^{dia} \tag{29}$$

Thus, as long as $\tilde{\lambda} \gg z\tilde{t}$ the system evolves so that $\Delta\rho^{dia} \sim e^{-(z^2/\tilde{\lambda})\tilde{t}}$, approaching zero (i.e., $\rho_{11}^{dia} = \rho_{22}^{dia} = 1/2$) at a rate $z^2/\tilde{\lambda}$ that decreases with increasing λ (Zeno effect). Obviously, for long enough time the opposite inequality $\tilde{\lambda} \ll z\tilde{t}$ will be realized. If $\tilde{\lambda}/z > 1$ this implies that at such times $\tilde{t} \gg 1$ or $V/ut \ll 1$, and no further population transfer takes place. If the system reached $\rho_{11}^{dia} = \rho_{22}^{dia} = 1/2$ before that time it will stay in this state. Otherwise evolution will freeze at some other value.[36] Both behaviors are seen in Figures 2, and in both cases the evolution is slower for larger $\tilde{\lambda}$, i.e., stronger measurement.

Finally, Figures 3a,b present an overall view of the behavior of the transition probability as function of the LZ parameter z and the measurement strength parameter λ. Both figures now show the probability to remain at $t \to \infty$ in the initial adiabatic state, however the evolutions are done under different observation conditions: Fig. 3a shows the results of evolution in which populations in the adiabatic levels are observed, while Fig. 3b shows the corresponding results when propagation is done under continuous observation of populations in the diabatic levels.

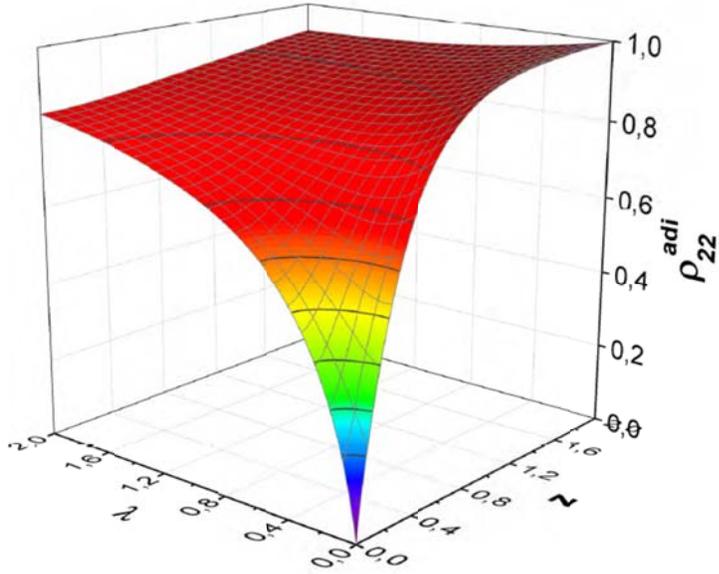

(a)

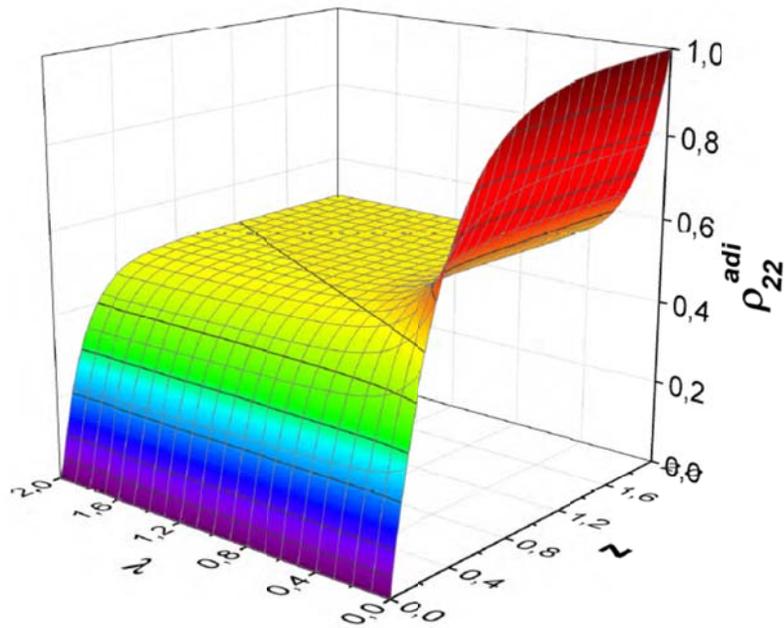

(b)

Fig. 3. Survival probability of the *adiabatic state*, $\rho_{22}(\tilde{t} \to \infty)$, starting from $\rho_{22}(\tilde{t} \to -\infty) = 1$, in a system undergoing continuing measurement of the adiabatic (a) and diabatic (b) populations, shown as functions of the Landau-Zener parameter $\tilde{z}$ and measurement strength $\tilde{\lambda}$. The asymptotic value is obtained at $\tilde{t} = 200$.

## 4. Discussion and conclusions

We have found that the time evolution associated with the Landau-Zener process under continuous weak population measurement depends on the character of the measurement process: When population of the adiabatic states is monitored, the time evolution exhibits a quantum Zeno effect behavior, becoming more adiabatic for stronger measurement. Interestingly, close to the adiabatic limit $((z > 1)$ the dependence on the measurement strength is non-monotonic, and adiabaticity initially decreases with increasing measurement strength, reflecting the effect of level broadening in this limit. When the population of the diabatic states is continuously detected, the Zeno effect is manifested in a slower time evolution under stronger measurement conditions, however, in contrast to the accepted notion concerning this effect, the asymptotic populations do not freeze in their initial values, but instead approach the value ½ (different asymptotic populations are realized if the energy levels separate before this value is reached).

These results should not be surprising in view of past work on the dynamics of the LZ process in a system interacting with a dissipative environment, [5-14] [15, 16] [17-21] including the possibility of externally affected control.[22] however viewed in the framework of measurement theory can yield some new insight. First is the strong dependence of the dynamics on the character of the measurement. Most of the papers cited above consider the effect on the LZ process of decoherence in the diabatic basis. In the present context, monitoring the population of the adiabatic states has a markedly different effect on the system dynamics than following the corresponding non-adiabatic states. Obviously, this difference just reflects the fact that environmental effects on system dynamics depend on the way the environment is coupled to the system, however viewed from the perspective of a measuring process this points to a way to controlling the system dynamics by engineering processes that affect its decoherence. [37, 38]

Secondly, the manifestation of the Zeno effect when the measurement is done in the non-adiabatic basis calls into question the standard measurement theory argument for this effect. This standard argument, applied to the dynamics of a two-coupled level system described by the Hamiltonian $\hat{H} = \begin{pmatrix} E_1 & 0 \\ 0 & E_2 \end{pmatrix} + \begin{pmatrix} 0 & V \\ V & 0 \end{pmatrix}$ and starting is state 1, points out that if, during a time interval $T$, $N$ projective measurements are done to determine whether the system is still in state 1, the probability to remain in this state at time $T$ is

$$P_1 \approx \frac{1}{2}\left(1 + \exp\left[\frac{-2V^2 T^2}{\hbar^2 N}\right]\right) \quad (30)$$

Which becomes 1 as $N \to \infty$. This argument, however, disregards the question whether projective measurements can be made at arbitrarily short time intervals, and arguments against this possibility were made.[39] Without getting into this discussion we note that the theory of continuous weak measurement implicitly assumes that the strength of individual measurements is inversely proportional to the measurements frequency, see Eq. (9). Indeed, it is easy to show that the argument that leads to Eq. (29) and consequently to $P_1 = 1/2$ in the strong measurement limit ($\lambda \to \infty$) of the dynamics described by Eqs.(30), remains valid when the functions $2ut$ on the r.h.s. of these equations are replaced by the constant $E_2 - E_1$, that is, when the LZ problem is reduced to the 2-coupled levels problem. Consistent with this is the observation that a similar result, $P_1 = 1/2$, is obtained as the $T \to \infty$ limit of Eq. (30) under the assumption that inverse frequency $\Delta t = T/N$ of projective measurements must be finite.

It is of interest to consider scenarios for experimental realization of such different measurements. Diabatic poulations can in principle be monitored in systems where the two diabatic states correspond to two molecular (or dot) charging states. It is harder to envisualize a measurement of adiabatic populations: The standard tool for such measurement is optical spectroscopy which is inherently a destructive measurement. Identifying a property of the eigenstates of a system's Hamiltonian that can be detected without destroying the state is an interesting challenge.

**Acknowledgements**. We are grateful to the Kurt-Lion-Foundation, the DFG-SFB 767 Controlled Nanosystems, and the EDEN (Erasmus Mundus Academic Network) for financial support. The research of AN is supported by the Israel Science Foundation, the Israel-US Binational Science Foundation (grant No. 2011509), and the European Science Council (FP7/ERC grant No. 226628).